\documentstyle[prl,aps,epsfig,multicol]{revtex}
\begin{document}
\draft
\title{The physical determinants of the thickness of lamellar polymer crystals}
\author{Jonathan P.~K.~Doye and Daan Frenkel}
\address{FOM Institute for Atomic and Molecular Physics, 
Kruislaan 407,\\ 1098 SJ Amsterdam, 
The Netherlands}
\date{\today}
\maketitle
\begin{abstract}
Based upon kinetic Monte Carlo simulations of crystallization in
a simple polymer model we present a new picture of the mechanism by which 
the thickness of lamellar polymer crystals is constrained to a value close 
to the minimum thermodynamically stable thickness.
This description contrasts with those given by the two dominant theoretical approaches.
\end{abstract}
\pacs{81.10.Aj,68.45.Da,65.60.Qb}

\begin{multicols}{2}

\begin{figure}
\begin{center}
\epsfig{figure=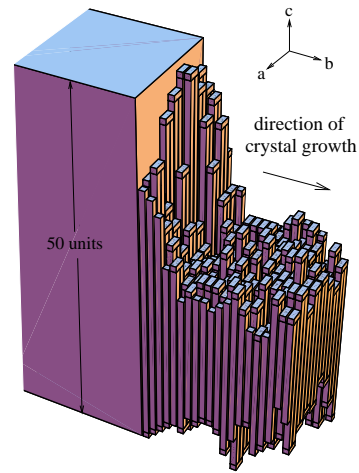,width=6.6cm}
\vglue -0.2cm
\begin{minipage}{8.5cm}
\caption{\label{fig:3D}Cut through a polymer crystal which was produced by the growth of 
twenty successive layers on a surface with a uniform thickness of 50 units at a temperature, 
$T=2.0\,\epsilon k^{-1}$.  
Folds occur at the top and bottom surfaces, and 
the stems are represented by vertical cuboids. The cut is 16 stems wide. 
}
\end{minipage}
\end{center}
\end{figure}

On crystallization from solution and the melt many polymers form lamellae
where the polymer chain traverses the thin dimension of the crystal many times folding back
on itself at each surface 
\cite{Keller57a}. 
(The crystal geometry is shown by the example configuration in Figure \ref{fig:3D}).
Although the lamellar crystals were first observed over forty years ago 
their physical origin is still controversial. 
It is agreed that the kinetics of crystallization are crucial since extended-chain crystals are
thermodynamically more stable than lamellae. 
However, the explanations for the dependence of the lamellar thickness on temperature  
offered by the two dominant theoretical approaches appear irreconcilable \cite{theorynote}.  
The lamellar thickness is always slightly greater than $l_{\rm min}$, the minimum thickness for which
the crystal is thermodynamically stable;
$l_{\rm min}$ is approximately inversely proportional to the supercooling.

The first theory, which was formulated by Lauritzen and Hoffman soon after the initial
discovery of the chain-folded crystals \cite{snuc},
invokes {\em surface nucleation} of a new layer on the thin side faces of the 
lamellae as the key process.
It assumes that there is an ensemble of crystals of different thickness 
each of which grows with constant thickness; 
the thickness for which the crystal has the maximum growth rate is then sought. 
The growth rates are derived by assuming that a new crystalline layer grows by the deposition
of a succession of stems (straight portions of the polymer that traverse the crystal once)
along the growth face. 
The two main factors that determine the growth rate are the thermodynamic driving force 
and the free energy barrier to deposition of the first stem in a layer. 
The former only favours crystallization when the thickness is greater than $l_{\rm min}$;
the latter increases with the thickness of the crystal because of the free energetic cost 
of creating the two new lateral surfaces on either side of the stem and makes 
crystallization of thick crystals prohibitively slow. 
Therefore, the growth rate passes through a maximum at an intermediate value of the thickness
which is slightly greater than $l_{\rm min}$.

The second approach which has been termed the {\em entropic barrier} model, 
is based upon the interpretation of kinetic Monte Carlo simulations \cite{Sadler84a,Spinner95}
and rate-theory calculations \cite{EBrate}
of a simplified model of the polymer crystal growth. As with the surface nucleation approach, 
the observed thickness is suggested to result from the competition between a driving force 
and a free energy barrier contribution to the growth rate. 
However, a different cause for the free energy barrier is postulated.
As the polymer surface in the model can be rough, 
it is concluded the details of surface nucleation of new layers are not important. 
Instead, the outer layer of the crystal is found to be thinner than in the bulk; 
this rounded crystal profile prevents further crystallization.
Therefore, growth of a new layer can only begin once a fluctuation occurs to an entropically
unlikely configuration in which the crystal profile is `squared-off'. 
As this fluctuation becomes more unlikely with increasing crystal thickness, 
the entropic barrier to crystallization increases with thickness.

Although both approaches are able to describe correctly some of the basic
phenomenology of polymer crystallization, both have questionable aspects.
For example, the surface nucleation assumption that crystals grow 

\vglue -0.3cm

\noindent
with constant thickness is contradicted by experiments in 
which temperature changes cause 
the thickness of growing lamellae to adjust to the new temperature \cite{Bassett62}.
And in the entropic barrier approach it is not clear whether approximations such as 
the implicit representation of the chain connectivity and chain folds 
by a set of simple growth rules compromise its conclusions.

The approach we take combines some of the more attractive features
of the two models.
As in the surface nucleation model, we form new polymer layers 
by the sequential growth of stems, and growth is considered to 
proceed layer by layer. However, we remove any constraints on 
the stem length. This additional complexity necessitates
the use of 
simulations to characterize the resulting behaviour.
Others have recognized the need for such a `multiple pathway'
approach, but, because of limited computational resources at that time,
only approximate treatments 
were possible\cite{Frank60,Lauritzen67b,Point79a,DiMarzio82a}.

We model the polymer by a self-avoiding walk on a simple cubic lattice. 
There is an attractive energy, -$\epsilon$, between non-bonded polymer units on 
adjacent lattice sites and between the polymer and the surface, 
and an energetic penalty, $\epsilon_g$, for kinks (`gauche bonds') in the chain. 
$\epsilon_g$ defines the stiffness of the chain.
Here, we use $\epsilon_g=8\epsilon$, 
however similar results are obtained at any positive $\epsilon_g$.
We only model the crystalline portion of the polymer explicitly and 
the rest of the chain is assumed to behave like a two-dimensional ideal coil
adsorbed onto the surface\cite{coilnote}.

At each step in our kinetic Monte Carlo simulation \cite{Voter86} changes of 
configuration can only occur at the ends of the crystalline portion of the polymer. 
The possible processes are for the crystalline layer to grow by one unit 
by the extension of one of the stems or by the creation of a fold, or for the crystalline
part of the polymer to shrink by one unit. 
The free energy change for a step
is given by $\Delta A=\Delta E_{\rm xtal}+\Delta A_{\rm coil}$, where
$\Delta E_{\rm xtal}$ is the change in energy of the crystalline configuration 
and $\Delta A_{\rm coil}$ is the change in free energy of the coil.
From this we derive a rate for a transition between the current state $i$ and the
state $j$: $k_{ij}={\rm min}\left( 1,\exp \left(-\Delta A_{ij}/kT\right) \right)$.
We then randomly choose one of the states connected to $i$, 
with probabilities given by $p_{ij}=k_{ij}/\sum_{j'}k_{ij'}$
and update the time by an increment $\Delta t= -\log(\rho)/\sum_{j'}k_{ij'}$,
where $\rho$ is a random number in the range [0,1]. 

In this way we can grow a new crystalline layer on a surface which represents
the growth face of the polymer crystal.
By taking averages over many such layers we can statistically characterize
the properties of the new layer.

\begin{figure}
\begin{center}
\epsfig{figure=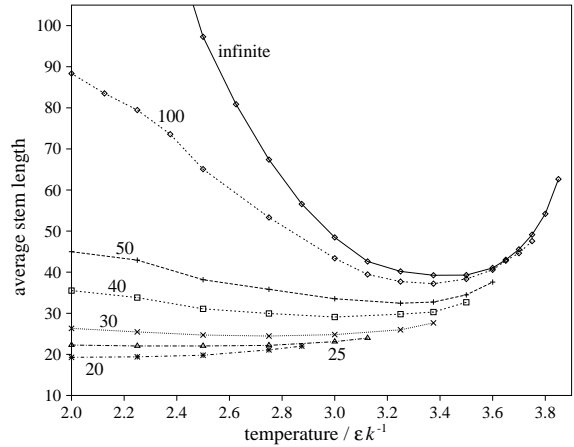,width=8.1cm}
\vglue 0.1cm
\begin{minipage}{8.5cm}
\caption{\label{fig:thick.T} The average stem length in the new crystalline layer 
as a function of temperature. 
The different curves are for different thicknesses of the underlying surface, as labelled.
}
\end{minipage}
\end{center}
\end{figure}

In Figure \ref{fig:thick.T} we show how the thickness of the new layer depends 
both on temperature and the thickness of the underlying layer. 
We immediately see that the thickness of the new layer is not necessarily the same
as that for the previous layer, showing that the constant thickness assumption of
the surface nucleation approach does not hold for our model. 
For example, for growth on a surface that is 100 units thick, the new layer is always thinner. 
As the temperature approaches the melting temperature ($T_m=4.06\,\epsilon k^{-1}$)
the thickness rises due to the increase 
of $l_{\rm min}$. At low temperature the thickness also
increases because it becomes harder to scale the free energy 
barrier for forming a fold, and
so on average the stem continues to grow for longer. 

\begin{figure}
\begin{center}
\epsfig{figure=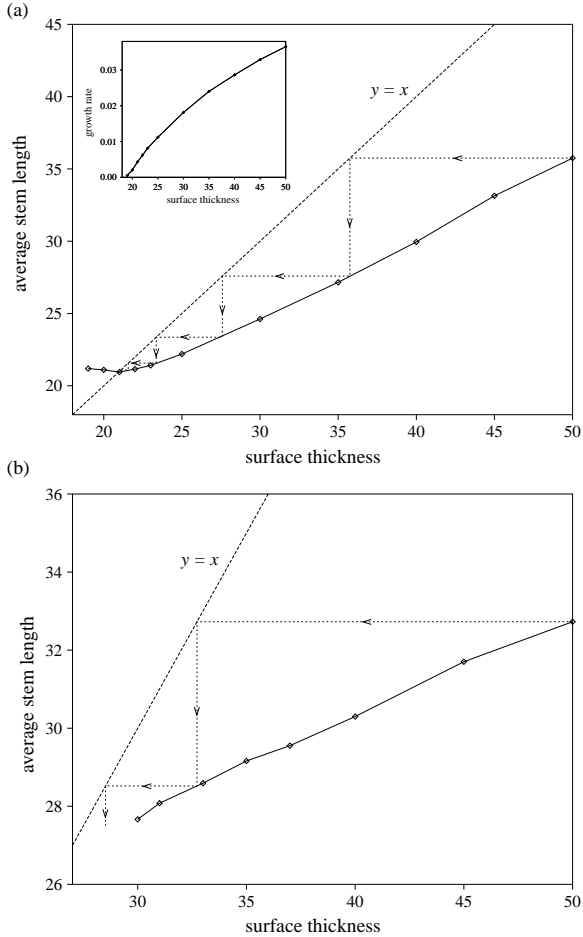,width=8.1cm}
\vglue 0.1cm
\begin{minipage}{8.5cm}
\caption{\label{fig:thick.l} 
The average stem length in the new crystalline layer 
as a function of the thickness of the underlying surface for 
(a) $T=2.75\,\epsilon k^{-1}$, (b) $T=3.375\,\epsilon k^{-1}$.
The dotted lines show how the thickness changes on addition of successive
layers to a 50-unit thick surface.
The inset of (a) shows the growth rate (polymer units per unit time) 
of the new layer as a function of thickness for $T=2.75\,\epsilon k^{-1}$. }
\end{minipage}
\end{center}
\end{figure}

A different perspective
is obtained by plotting 
the thickness of the new layer against the thickness of the underlying surface, 
as in Figure \ref{fig:thick.l}.
It also allows us to consider multi-layer growth implicitly.
At $T=2.75\,\epsilon k^{-1}$ growth on a 50-unit thick surface produces a new layer
of thickness $\sim$36. Growth of another layer on top of this layer is 
similar to growing a layer on a surface with a uniform thickness of 36; the latter
gives a new layer of thickness $\sim$28. 
We can see the effect of adding each successive layer
by following the dotted line in Figure \ref{fig:thick.l}a; 
it also clearly shows that the mapping is a fixed-point attractor.
The thickness of the crystal converges to the thickness at which the curve crosses
$y=x$, i.e.\ to the point where the thickness of the new layer is the same as the previous.
Therefore, at this temperature, there is only one thickness for which stable growth occurs
(the inset of Figure \ref{fig:thick.l}a shows that this is not the thickness at which growth of a new 
layer is the most rapid)
and there is no barrier to growth due to a rounded crystal profile.

Although our results contrast with the assumptions of both surface nucleation 
and entropic barrier models that the observed thickness corresponds to the 
one with the maximum growth rate, in a number  of the early polymer crystallization 
papers\cite{Frank60,Lauritzen67b,Price61} it was
realized that stable growth could only occur at the thickness for which
a new layer has the same thickness as the previous.
However, since then this insight has for the most part been neglected,
and, to the best of our knowledge, the iterative maps that underlie it 
have not been previously visualized.

The above picture is confirmed when we simulate actual multi-layer growth.
The cut through a crystal depicted in Figure \ref{fig:3D} shows the crystal 
converging to the stable thickness within 5--10 layers and then
continuing to grow at that thickness. 

\begin{figure}
\begin{center}
\epsfig{figure=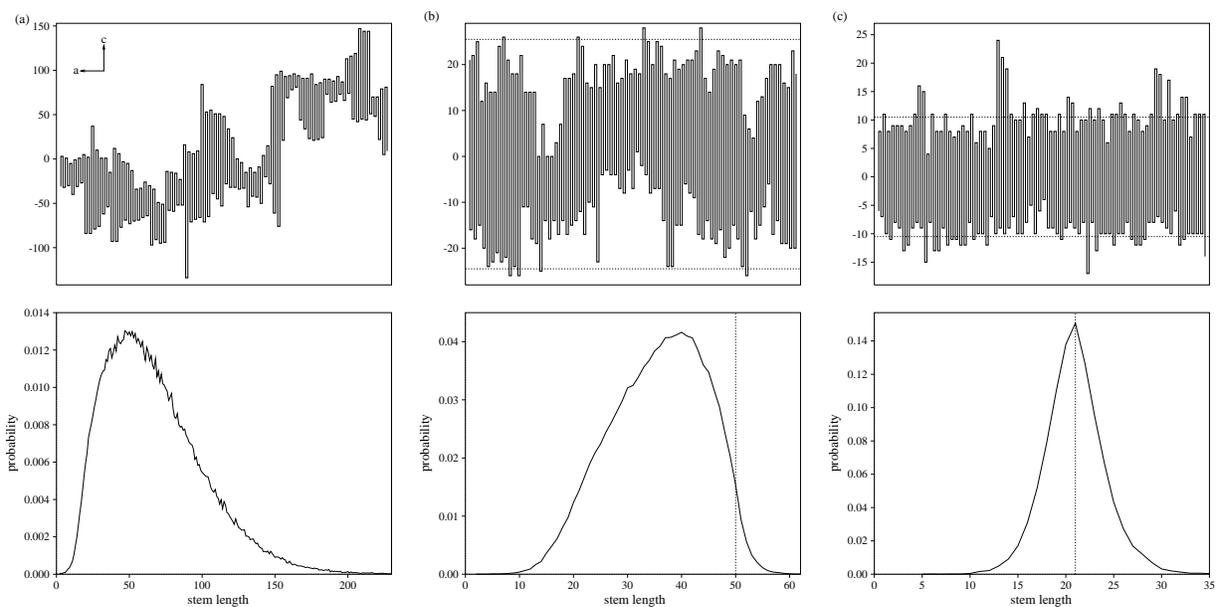,width=16.3cm}
\vglue0.1cm
\begin{minipage}{17.6cm}
\caption{\label{fig:config}Polymer configurations and probability distributions
of the stem length for a new crystalline layer grown at $T=2.75\,\epsilon k^{-1}$ 
on a surface of thickness: (a) $\infty$ (b) 50 and (c) 21. 
Each configuration contains 150 stems and a dashed line marks the boundaries of 
the underlying surface; the $y$-axis origin is at the centre of the surface.
The dashed vertical line in the probability distributions in (b) and (c) is
at the thickness of the underlying surface.}
\end{minipage}
\end{center}
\end{figure}

We can understand better the reasons for this behaviour by examining example 
polymer configurations and the probability distributions of the stem length
for the growth of a single layer at a number of surface thicknesses (Figure \ref{fig:config}).
$l_{\rm min}$ places one constraint on the stem length; only a small
fraction of the stems can be shorter than $l_{\rm min}$ 
if the layer is to be thermodynamically stable. 
The boundary of the growth face places the second constraint on the stem length; 
it is energetically unfavourable for the polymer to extend beyond the edges of the underlying surface. 
Even in the absence of this constraint, 
i.e.\ on the infinite surface, the stem length remains finite because at every step 
there is always a finite probability that a fold will be formed \cite{Point79a,DiMarzio82a}, 
and so the probability distribution decays to zero at large stem lengths (Figure \ref{fig:config}a).

The effect of the finite thickness of the growth face can be clearly seen 
in the growth of a layer on a 50-unit thick surface (Figure \ref{fig:config}b); 
the probability of the stem length being greater than the surface thickness is much less
than it being smaller. Therefore, the new layer is thinner than the underlying surface. 
Only when the surface thickness approaches $l_{\rm min}$ does the probability distribution
become symmetrical about the surface thickness and the thickness of the new layer become 
equal to the surface (Figure \ref{fig:config}c). 
This is the reason why the thickness at which stable growth occurs is close to $l_{\rm min}$.

The above scenario does not hold for all temperatures.
The last point of the curve in Figure \ref{fig:thick.l}a is for a surface 19 units thick; 
a new layer cannot grow on a thinner surface. 
However, there is no {\it a priori\/} reason why the
thickness curve must end after the curve has crossed $y=x$.
Indeed, for $T>3.2\,\epsilon k^{-1}$ there is no thickness for which successive layers
have the same thickness. 
For example, after the growth of two layers on a 50-unit thick surface 
the outer layer is $\sim$29 units thick (Figure \ref{fig:thick.l}b); the crystal 
then stops growing because the outer layer is too thin.
For these smaller supercoolings, as suggested in the entropic
barrier model, the entropic rounding of the crystal profile inhibits growth.

To overcome this entropic barrier requires a cooperative mechanism
whereby a new layer takes advantage of (and then locks in) 
dynamic fluctuations in the outer layer to larger thickness.
The presence of such fluctuations is shown in the probability distributions of 
Figure \ref{fig:config};
however, growth stops in our model because we attempt 
to grow a new layer on an outer layer that is static.
As overcoming the barrier would be most rapid when the magnitude of the fluctuations is the minimum necessary,
we expect that this mechanism would lead to the crystal continuing to grow 
with a thickness that corresponds to the smallest for which a new layer can grow, 
e.g. at $T=3.375\,\epsilon k^{-1}$ this is a thickness of 30 (Figure \ref{fig:thick.l}b).
This mechanism again leads to a thickness for the polymer crystal which is close to $l_{\rm min}$.

In summary, our results present a new picture of the mechanisms that 
cause the lamellar thickness to be constrained to a value just above
$l_{\rm min}$. 
The free energetic costs of the polymer extending beyond the edges of 
the previous crystalline layer and of a stem being shorter than $l_{\rm min}$
provide upper and lower constraints on the length of stems in a new layer.
Their combined effect is to cause
the crystal thickness to converge to a value close to $l_{\rm min}$ where
growth with constant thickness then occurs.

This convergence of the thickness has been observed in 
experiments in which the thickness of growing
polymer crystals adjusts to a change in temperature \cite{Bassett62} and in 
\end{multicols}
\begin{multicols}{2}
\noindent
which lamellae form by epitaxial crystallization onto extended-chain 
crystalline fibres \cite{Keller79}. It would be very 
interesting if atomic force microscopy (AFM) could
be used to probe the profiles of the steps on the lamellae that result from temperature changes; 
from these profiles maps similar to Figure \ref{fig:thick.l}a 
could be constructed. 
AFM could be also used to study the profile of the crystal close to the growth face
to examine whether entropic rounding of the crystal edge occurs.

Our description of polymer crystallization differs significantly from that given by 
the surface nucleation approach. 
The assumption that a crystal of any thickness can continue to
grow at that thickness does not hold. 
Furthermore, although  overcoming the free energy barrier to the 
nucleation of the first few stems in a new layer can be one of the slowest
processes in the growth of a new layer, it does not play a role in 
determining the thickness at which stable growth can occur.
By contrast, some of the insights of the entropic barrier model form 
part of the more complete picture we present here.

The work of the FOM Institute is part of the research program of
`Stichting Fundamenteel Onderzoek der Materie' (FOM)
and is supported by NWO 
(`Nederlandse Organisatie voor Wetenschappelijk Onderzoek'). 
JPKD acknowledges the financial support provided by 
the Computational Materials Science program of the NWO.

\end{multicols}
\end{document}